\documentstyle[12pt,aaspp4]{article}
\def\ie{i.e.\ }
\def\eg{e.g.,\ }
\def\etal{et~al.\ }
\def\ltsima{$\; \buildrel < \over \sim \;$}
\def\simlt{\lower.5ex\hbox{\ltsima}}
\def\gtsima{$\; \buildrel > \over \sim \;$}
\def\simgt{\lower.5ex\hbox{\gtsima}}

\def\kms{km s$^{-1}$}

\journalid{}{}
\articleid{}{}
\lefthead{Mihos, McGaugh, \& de Blok}
\righthead{Stability of LSB Disks}
\slugcomment{To appear in the {\it Astrophysical Journal Letters}}
 
\begin{document}
 
\title{Dynamical Stability and Environmental Influences in Low Surface 
Brightness Disk Galaxies}
 
\author{J. Christopher Mihos,\altaffilmark{1,2}
Stacy S. McGaugh\altaffilmark{3}, and W.J.G. de Blok\altaffilmark{4}}
 
\altaffiltext{1}{Hubble Fellow}
\altaffiltext{2}{Department of Physics and Astronomy, Johns Hopkins 
	University, Baltimore, MD 21218; hos@pha.jhu.edu}
\altaffiltext{3}{Carnegie Institute of Washington, Department of 
	Terrestrial Magnetism, 5241 Broad Branch Road, NW,
         Washington, DC 20015; ssm@dtm.ciw.edu}
\altaffiltext{4}{Kapteyn Astronomical Institute, P.O. Box 800,
         9700 AV Groningen, The Netherlands;
         blok@astro.rug.nl}

\begin{abstract}

Using analytic stability criteria, we demonstrate that, due to their low 
surface mass density and large dark matter content, LSB disks are quite 
stable against the growth of global nonaxisymmetric modes such as bars. 
However, depending on their (poorly constrained) stellar velocity dispersions, 
they may be only marginally stable against local instabilities. We simulate a 
collision between an LSB and HSB galaxy and find that, while the HSB galaxy 
forms a strong bar, the response of the LSB disk is milder, manifesting weaker 
rings and spiral features. The lack of sufficient disk self-gravity to amplify 
dynamical instabilities naturally explains the rarity of bars in LSB disks.  
The stability of LSB disks may also inhibit interaction-driven gas inflow and 
starburst activity in these galaxies.

\end{abstract}
 
 
\vfil\eject
 
\section{Introduction}

Low surface brightness (LSB) disk galaxies represent a common
product of galaxy formation and
evolutionary processes. Recent surveys have uncovered large numbers
of LSB galaxies (\eg Schombert \etal 1992) whose
central surface brightnesses are an order of magnitude fainter than
in their high surface brightness (HSB) counterparts. These LSB galaxies
can rival HSB galaxies in terms of size and luminosity,
and are estimated to contain as much as one-third of the total mass in
galaxies in the local universe (McGaugh 1996). They are rich
in HI and are forming stars at a very low rate, leading to the suggestion
that LSB galaxies represent relatively unevolved systems
(McGaugh \& Bothun 1994;  de Blok \etal 1995) whose
surface densities are too low to foster the dynamical processes
which may lead to star forming activity (van der Hulst \etal 1993).
 
The paucity of star formation in LSB disks may also be linked to their
local environment.  Although embedded in the same large scale structure as
HSB galaxies, LSB galaxies are less clustered on all scales (Mo \etal 1994)
and are particularly isolated on scales smaller than a few Mpc (Zaritsky
\& Lorrimar 1992; Bothun \etal 1993). Without the well-established
dynamical trigger provided by an interacting companion, LSB galaxies may
simply evolve passively due to their low surface densities, and never
experience any strong star-forming era in their lifetimes. Indeed,
sufficient tidally induced star formation in LSB disks may drive 
evolution from LSB to HSB galaxies. This has been suggested as the cause of the
observed isolation of LSB galaxies: interactions in denser environments
transform them into HSB galaxies or perhaps even destroy them entirely.
 
Dynamical modeling has shown that the ability for interactions to trigger
strong starbursts is linked to the onset of bar instabilities in the stellar
disk (\eg Noguchi 1987; Barnes \& Hernquist 1991; Mihos \etal 1992; Mihos
\& Hernquist 1994ab, 1996) -- instabilities which are largely governed
by the internal structure of disk galaxies. However, these 
efforts employed model galaxies constructed to match
typical HSB galaxies, and recent work has shown that the structural
properties of LSB galaxies are significantly different 
(de Blok \& McGaugh 1996; hereafter dBM). In particular, LSB galaxies have
lower disk mass densities, more slowly rising rotation curves, and a
higher fraction of dark to visible matter than do HSB galaxies.  As a
result, the strong bar-induced inflows and starburst activity manifested
in numerical simulations and observed in bright nearby interacting
systems may not typify the response of LSB galaxies to a gravitational
interaction.
 
The degree of stability of LSB disks could determine to a great extent
exactly how any collisionally-induced evolution would occur. Bars are
rare in LSB galaxies; of the 36 late type LSB galaxies in the studies of
McGaugh \& Bothun (1995) and de Blok \etal (1995), only one (F577-V1)
shows a strong bar. Similarly, a perusal of the LSB catalog by Impey
\etal (1996) shows only $\sim$ 20 barred systems out of a sample of
$\sim$ 500 LSBs. This frequency of bars is much lower than the $\sim$
30\% found for the field HSB galaxies which make up the RC2 catalog
(Elmegreen, Elmegreen, \& Bellin 1990). It is unclear, however, if
the scarcity of barred LSBs is due to their isolation or to a higher
degree of disk stability in LSB disks.

In this letter, we explore the dynamical stability of LSB disks in order
to judge their response to external perturbations. Both analytic stability
criteria and numerical simulations are used. We find that LSB disks are 
more stable against bar formation than their HSB counterparts, but that 
collisions may drive local instabilities in LSB disks. We close
with a discussion of our results in the context of collisionally
induced galaxy evolution.

\section{Structure of LSB Disks}

The rotation curves of LSB disks exhibit two general characteristics.
The first is a rigorous adherence to the Tully-Fisher relation (Sprayberry 
\etal 1995; Zwaan \etal 1995; Hoffman \etal 1996) so that galaxies of the same
luminosity have the same rotation velocity in the flat, outer part of the
rotation curve.  The second is a dependence of the rate of rise of the inner
part of the rotation curve on the surface brightness (de Blok \etal 1996).
This occurs in the obvious sense, with more diffuse galaxies having more
gradually rising rotation curves.  This difference becomes much less
pronounced when the radius is normalized by the scale length~--- $V(R/h)$
is generally quite similar at a given luminosity (\eg Persic \& Salucci
1996), though some difference often remains.
 
These properties have a striking consequence: lower surface brightness 
galaxies are progressively more dark matter dominated (de Blok \etal 1996, 
de Blok \& McGaugh 1997). Rotation curve decompositions indicate that
the decreased surface brightness in LSB disks is accompanied by a decrease
in mass surface density of the disk.  Since it is the disk self gravity 
which drives instabilities, the dark matter domination of 
LSB galaxies may lead to important differences in their response to 
perturbations.
 
Here we examine models motivated by two galaxies with comparable luminosities
but surface brightnesses differing by nearly 3 magnitudes: NGC 2403 and UGC 128
(dBM). The rotation curve for UGC 128 is typical of
rotation curves for a large sample of LSB disks by de Blok \etal (1996; see
their Figure 7). We decompose the rotation curves into disk and 
halo components assuming ``maximum disk,'' where we attribute as much mass 
to the stellar disk as allowed by the rotation curves. Under this assumption,
the disk surface density of UGC 128 is nearly an order of magnitude lower than 
that of NGC 2403, roughly comparable to the 2.8 B magnitude difference in 
central surface brightness.  With the shorter radial scale 
length for NGC 2403 (2.1 kpc vs. 6.8 kpc for UGC 128), the two galaxies have
very similar total disk masses.

It is far from obvious that the maximum disk solution is plausible in LSB
galaxies (see de Blok \& McGaugh 1997 for an extensive discussion).  In
the case of UGC 128, maximum disk gives $M/L_B = 3$ for the stars, 2 or 3
times what is reasonable from the standpoint of stellar populations
(McGaugh \& Bothun 1994; de Blok \etal 1995).  Nonetheless, we retain
the maximum disk solution because it is a conservative starting point for
our analysis:  if the actual disks are less massive than we assume, they
will be even more stable than we find.

\section{Analytic Stability Criteria}
 
Given the rotation curve decomposition from dBM (see their Table 1), we can 
now calculate a variety of disk stability parameters to examine
quantitatively the issue of LSB disk stability. Although the rotation
curves are not well constrained at large radius (\ie at R $>$ 5 scale lengths),
it is the stability and mass distribution in the inner few scale lengths
in which we are ultimately interested. LSBs are predominantly stellar
in this region even though they can have quite high gas contents at
larger radii. Uncertainties in the mass 
distribution at large radius will have little effect on the evolution
of the inner disk. The rotation curves of NGC 2403 and UGC 128 are shown 
in Figure 1a.

\begin{figure}[b!] 
\plotfiddle{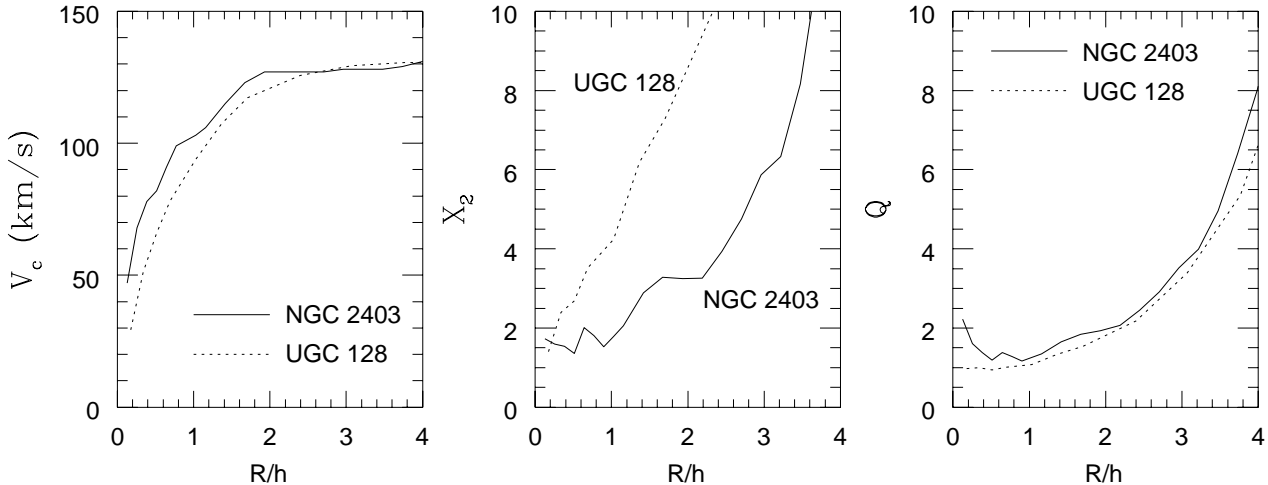}{2.0in}{270}{70}{70}{-300}{200}
\vspace{10pt}
\caption{Left: Rotation curves of NGC 2403 (HSB) and UGC 128 (LSB),
as a function of disk scale length ($R/h$).  Middle:
$X_2$ stability parameter. Right: Toomre $Q$ parameter. The two curves for
UGC 128 reflect two choices for $\sigma_r$.}
\label{fig1}
\end{figure}

Many criteria have been proposed to describe the global stability of
disk galaxies (\eg Ostriker \& Peebles 1973; Efstathiou \etal 1982;
Christodoulou \etal 1995).  Regardless of the
details of implementation, the fundamental argument behind these varying
criteria is that the presence of a massive dark matter halo helps to stabilize 
galaxies against bar formation (as long as the dark matter is dynamically hot). 
LSB disks, with their very low surface mass density and high dark matter
contents (dBM; de Blok \& McGaugh 1997), should be much more stable than their
HSB counterparts.

One measure of the susceptibility of galactic disks to global nonaxisymmetric
instabilities is the $X$ parameter (\eg Goldreich \& Tremaine 1978, 1979; 
Toomre 1981):
$$X_m \equiv {{k_{crit}R}\over{m}} = {{\kappa^2 R}\over{2\pi m G \Sigma_d}},$$
where $k_{crit}$ is the wavenumber of marginal stability, $\kappa$ is the 
epicyclic frequency, $R$ is the radius, $\Sigma_d$ is the disk surface 
density, and $m=2$ for bar modes. Toomre (1981)
showed that for flat rotation curves, disks proved stable against growing
modes if $X>3$, while for linearly rising rotation curves $X>1$ is 
a sufficient condition for stability (A. Toomre, private communication).
Nonetheless, galaxies with slowly rising rotation curves are believed
to be prone to bar formation due to two effects: their lowered
epicyclic frequency reduces $X$, and their large region of solid body
rotation suggests that, once formed, bars may exist for many dynamical 
times (Lynden-Bell 1979). The slowly rising rotation curves of LSB disks 
thus suggests
that they may be susceptible to the growth of global modes in the disk. 
However, their lowered mass surface density works in the opposite sense, 
and it is the competition between these two effects which determines their 
overall stability.

Given the rotation curve decomposition and mass modeling from dBM, we can 
calculate the $X_2$ parameter for NGC 2403 and
UGC 128 (note that the $m=2$ bar mode is the strongest non-axisymmetric 
instability in most disk galaxies). Figure 1b shows $X_2$ as a function
of radial scale length for these two galaxies. The high surface brightness
galaxy NGC 2403 is only marginally stable over a large range of radius:
in the inner regions $X \sim 1.5 - 2$, while at two scale lengths, where
the rotation curve has flattened, $X\sim 3$, still close to instability.
By contrast, the low surface brightness galaxy UGC 128 proves stable
throughout the disk, due to its lower mass surface density. We emphasize
that we have used maximum disk models; if LSBs are less than maximal disks, 
they will be even {\it more} stable. 

If LSB disks are stable against the growth of global instabilities in the 
disks, are they also stable against {\it local} instabilities? The growth
of local axisymmetric instabilities is measured by Toomre's $Q$ parameter 
(Toomre 1964),
$$Q\equiv { {\sigma_r \kappa} \over {3.36 G \Sigma_d} },$$ 
where $\sigma_r$ is the radial velocity dispersion of the disk stars. The 
determination of $Q$ is more problematic than $X_2$, because of the explicit 
dependence on $\sigma_r$. Velocity dispersions in LSB disks have not been 
directly measured, due to their very low surface brightness nature.  If LSB 
disks have velocity dispersions comparable to HSB disks, they will be quite 
stable due to their lowered disk surface mass density. Alternatively, if 
stellar velocity dispersions are linked to mass surface density, $Q$ for 
HSB and LSB disks may be similar. In this case, LSBs are globally stable 
against bars, but may only be marginally stable to local perturbations. 
For example, Figure 1c shows Q in each disk, assuming $\sigma_r=30$ 
\kms\ for NGC 2403 (similar to the value in the solar neighborhood; Mihalas 
\& Binney 1981)  and that velocity dispersion and surface density follow the
relation $\sigma^2 \sim \Sigma_{d}$ as might be expected from simple energy
arguments. While velocity dispersion most likely varies with radius,
these arguments suggest that LSB and HSB disks may have similar 
{\it local} stability properties, and that local instabilities might grow 
in LSB disks where global modes cannot. This result is is similar to
that of van der Hulst \etal (1993), who found the gas in LSB galaxies 
was only marginally stable against local instabilities. 

\section{Dynamical Modeling}

The analytic arguments of \S 3 indicate that LSB disks will be quite
stable against bar formation, but perhaps susceptible to local 
instabilities in the disk. These arguments are based largely on linear 
perturbation theory, however, and it is unclear just how LSB disks would 
respond to a strong perturbation such as an interaction with a neighboring
galaxy.  To examine this situation, we use numerical simulation to model a 
grazing encounter between an LSB galaxy and an HSB companion. We choose a 
zero-energy parabolic orbit, with a perigalactic separation of $R_p=10$ disk 
scale lengths. The collision is perfectly prograde, maximizing the tidal 
effects acting on the galaxy disks.

Rather than build galaxy models which differ in a number of structural
parameters, we focus on variations in disk surface density to define
the difference between HSB and LSB disk galaxies. We begin with a fiducial
composite disk/halo galaxy model for the HSB galaxy, built as described by
Hernquist (1993).  This model consists of a stellar disk of mass $M_d=1$ and 
exponential scale length $h=1$, and a truncated isothermal dark matter
halo of mass $M_h=5.8$, core radius $\gamma=1$, and exponential cutoff radius
$r_c=10$. In these units, the (disk) half-mass rotation period is 13 time
units. To construct the LSB galaxy, we use the same set of structural
parameters, except that the disk mass is chosen to be $M_d=1/8$, and scale
the galaxies to have identical {\it total} mass. 
Lacking information on the observed stellar velocity dispersion 
in LSB disks, we initialize velocities in {\it both} galaxy disks such that
Q=1.5. This conservative assumption implies lower velocity dispersion
in the LSB disk, as might be expected if disk surface density determines
the stellar velocity dispersion. If velocity dispersions in LSB disks are 
comparable  to those in HSB disks, LSB galaxies will be have a higher Q and
be more stable than our models. Our simulation is thus a conservative test of 
LSB stability. In the stellar dynamical models shown here, each halo is
represented by 131,072 particles, while the disks are comprised of 32,768
particles each. The large number of halo particles used minimizes that
the growth of instabilities due to discreteness noise in the halos 
(see, \eg Walker, Mihos, \& Hernquist 1996).

\begin{figure}[b!] 
\plotfiddle{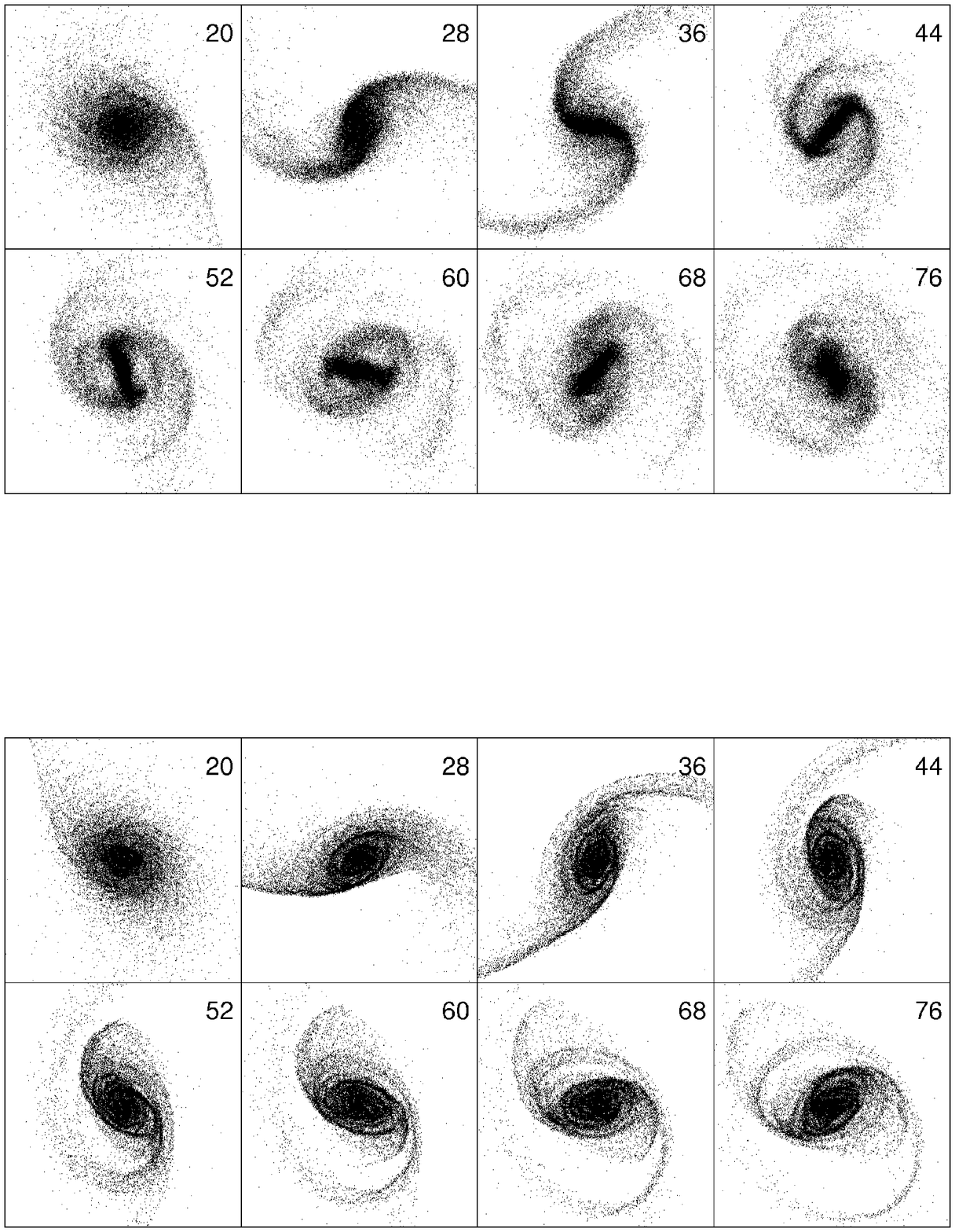}{7in}{0}{65}{65}{-200}{0}
\caption{Evolution of the HSB (top) and LSB (bottom) disk during an
equal (total) mass galaxy encounter with $R_{peri}=10$ scale lengths. Each frame
measure 7.5 scale lengths on a side, and time is given in the upper right
corner. One half-mass rotation period is roughly 13 time units.}
\label{fig2}
\end{figure}

Figure 2 (Plate X) shows the evolution of the disks in the HSB-LSB interaction, 
viewed in the orbital plane. Closest approach occurs at T=24, and the
galaxies respond quite strongly to the interaction, showing oval distortions
and tidal arms shortly after the encounter. In the HSB disk, the self-gravity
of the disk amplifies the perturbation such that by T=44 the galaxy has 
developed a very strong bar. This bar persists to the end of the simulation 
(T=80), significantly heating the inner disk. By contrast, the response
of the LSB disk is milder, although still quite significant. The encounter
strongly perturbs the galaxy, but without adequate self-gravity in the disk,
no bar develops during the simulation. Instead, the LSB disk displays 
long-lived spiral arms and rings in the disk, and a persistent oval
distortion. The crispness of the features in the LSB disk are attributable
to the low velocity dispersion corresponding to Q=1.5. If, instead, the
velocity dispersion in LSB disks is comparable to that in HSB disks, the
sharpness of these features will be reduced, similar to those in the simulated
HSB disk.

To quantify the response of the galaxies to the interaction, Figure 3 shows
a Fourier analysis of the growth of the $m=2$ mode in the inner half mass of 
each disk. The LSB galaxy actually responds first; due to its shallower 
potential well, the disk is more easily distorted by the
approaching companion. After the encounter, the LSB disk settles into
equilibrium, with $A_2$ relatively constant thereafter.\footnote{$A_2$ 
is the amplitude of the $m=2$ Fourier coefficient; see Sellwood \& Athanassoula
(1986).} Because of the high dark matter content of the LSB disk galaxy,
any subsequent evolution in the $m=2$ component will occur very slowly,
if at all.  Meanwhile, the growth of the $m=2$ mode in the HSB galaxy is more 
dramatic, as the bar continues to grow due to the disk self-gravity. The peak 
strength in $A_2$ is more than twice that of the LSB disk, and declines at 
late time, probably due to disk heating by the bar. We emphasize that
the $m=2$ mode is not only different in strength between the disks,
but also in character: the HSB sports a strong bar, while the LSB displays
a milder oval distortion. The influence of these features on the
structural and hydrodynamic evolution of the galaxies will be quite
different.

\begin{figure}[b!] 
\plotfiddle{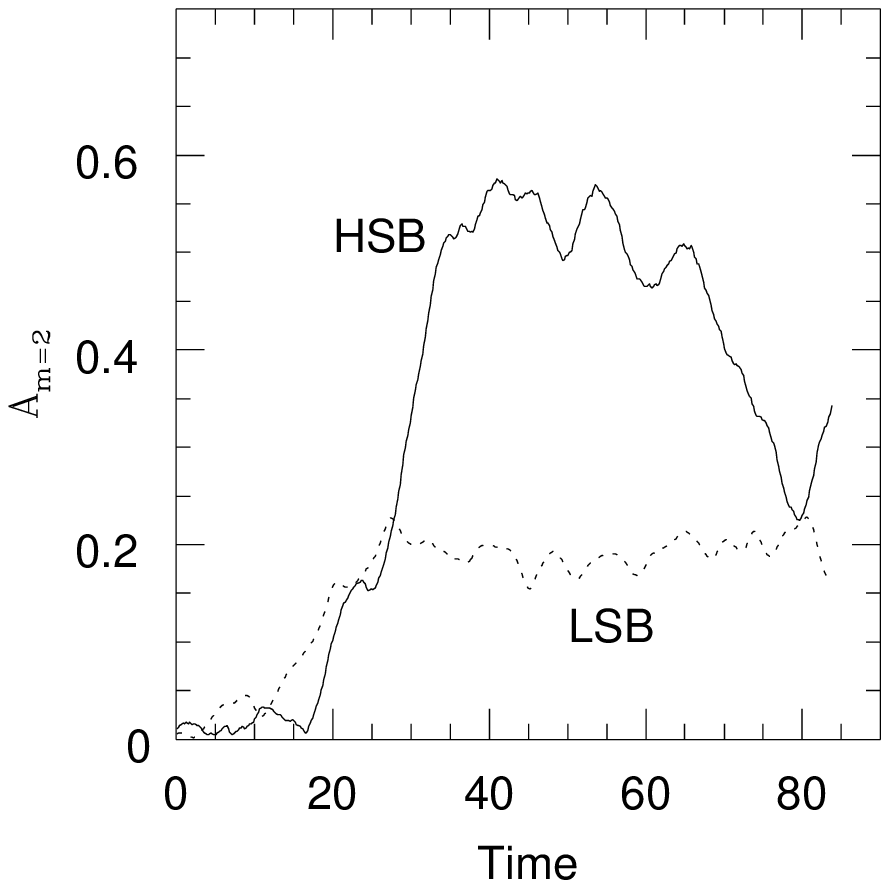}{2in}{0}{100}{100}{-200}{-400}
\caption{Growth of $m=2$ modes in the HSB and LSB disk during the
encounter.  Closest approach occurs at $T=24$.}
\label{fig3}
\end{figure}

\section{Discussion}

Both analytic arguments and numerical simulation indicate that, despite
their seemingly fragile nature, LSB disks are quite stable against the
growth of bar instabilities. Their low disk surface density and high dark 
matter content deprives LSB galaxies of the disk self-gravity necessary to 
amplify any nonaxisymmetric dynamical seeds. The stability of these galaxies
helps explain the rarity of observed bars in LSB disks. 

The stability of LSB disks also has interesting ramifications for scenarios
involving LSB galaxy evolution. Although the simulations reported here are 
purely stellar-dynamical, preliminary models which include hydrodynamics show 
that without the driving force of a bar, there is no strong inflow of gas 
to the galaxy center. This is a problem for the otherwise
appealing notion that LSB dwarf galaxies are the progenitors of
\ion{H}{2} galaxies experiencing central starbursts (Taylor \etal 1994),
if tidal encounters drive gas inflow. 
The need for an LSB progenitor population (McGaugh 1996b) and the similarity
of the environments of LSB and \ion{H}{2} galaxies (Salzer 1989;
Telles \& Terlevich 1995) strongly suggests such a connection, but even the
relatively close, strong interaction we have presented will not result in a 
strong central starburst. In order to provoke a violent enough response
in the LSB disk, a bona-fide merger may be necessary.
Furthermore, we have modeled rather large galaxies;
the lower luminosity galaxies which are potential \ion{H}{2} galaxy
progenitors are even more dark matter dominated than our models, making it
extremely difficult to drive the instabilities and radial inflows which
trigger central starbursts.

Another proposed evolutionary scenario appeals to the fragility of
LSB disks as an explanation for the low density environments in
which these galaxies are found. LSB disks avoid high
density regions, and are particularly isolated on small ($< 2$ Mpc)
scales (Bothun \etal 1993, Mo \etal 1994).  The fragile appearance of
LSB disks has lead to the idea that LSBs in dense environments might be 
destroyed or structurally altered beyond recognition, or that star formation 
induced by tidal encounters may transform LSBs into HSBs.
The dynamical arguments presented here indicate
that LSBs are sufficiently stable to survive galaxy encounters structurally
intact. However, while LSBs are robust against global instabilities, the sharp 
spiral features which form may correspond to {\it local} compressions which 
could push disk gas above some critical threshold and ignite star formation 
throughout the disk, and perhaps lead to surface brightness evolution which 
transforms LSBs into HSB galaxies.
 
A number of considerations, however, argue against the transformation
of LSBs into HSBs via tidal encounters. Having modeled a strong interaction, 
it is not certain that more distant encounters can drive any significant 
perturbations. While mergers may drive more dramatic evolution, the isolation 
of LSBs occurs on large scales, apparent out to at least 1 Mpc.  It is hard to 
understand how such stable galaxies can be affected at such enormous ranges 
since the severity of encounters declines strongly as the impact parameter 
increases.  Moreover, any global enhancement of the surface brightness can 
not be so severe as to remove a galaxy from the Tully-Fisher relation, so the
evolution hypothesized to explain the lack of LSBs in dense regions
is probably not sufficient to fully transform their identity
from LSB to HSB.  Interactions may neither destroy these late type
systems (as advocated by Moore \etal 1996), nor transform them
structurally into systems like the Milky Way.
 
We are thus left with a serious conundrum.  The isolation of LSB galaxies
and their potential role as the progenitors of \ion{H}{2} galaxies is
qualitatively well explained by tidally induced star formation.  Quantitative
analysis suggests that the dark matter domination of LSB disks makes 
these fragile-looking galaxies
surprisingly stable and resistant to this process.  It may thus be that
isolation is a prerequisite for the {\it formation\/} of LSB galaxies.  Yet the
environmental connection remains compelling and warrants further investigation.
\acknowledgements

We thank Alar Toomre and Scott Tremaine for valuable discussions on disk 
stability issues.  This work was sponsored in part by the San Diego 
Supercomputing Center.  J.C.M. is supported by NASA through a Hubble 
Fellowship grant \#~HF-01074.01-94A awarded by the Space Telescope Science 
Institute, which is operated by the Association of University for Research in
Astronomy, Inc., for NASA under contract NAS 5-26555.

\end{document}